\documentclass[11pt]{article}

\usepackage{fullpage, hyperref,bbm,amssymb,amsfonts,amsthm}
\usepackage[noend]{algorithmic}
\usepackage{algorithm}
\usepackage{setspace}
\usepackage{color}
\usepackage{latexsym}
\usepackage{verbatim} 
\usepackage{graphics}
\usepackage{graphicx}

\input{Qcircuit}

\renewcommand{\>}{\rangle}

\newtheorem{theorem}{Theorem}

\newtheorem*{problem}{Problem}

\begin{document}
\title{Selecting Efficient Phase Estimation With Constant-Precision Phase Shift Operators} 
\author{ 
 		Chen-Fu Chiang \thanks{ D\'{e}partement de Physique, Universit\'{e} de Sherbrooke Sherbrooke,
		Qu\'{e}bec, Canada J1K 2R1S. \newline \qquad Email:{{Chen-Fu.Chiang}@USherbrooke.ca}} 
	}		 
		
\maketitle

\begin{abstract}
We investigate the cost of three phase estimation procedures that require only
constant-precision phase shift operators. The cost is in terms of the number of elementary gates,
not just the number of measurements. Faster phase estimation requires the minimal number
of measurements with a $\log *$ factor of reduction when the required precision $n$ is large. 
The arbitrary constant-precision approach (ACPA) requires the minimal number of elementary gates
with a minimal factor of 14 of reduction in comparison to Kitaev's approach. The reduction 
factor increases as the precision gets higher in ACPA. Kitaev's approach
is with a reduction factor of 14 in comparison to the faster phase estimation in terms of 
elementary gate counts.
\end{abstract}
%%%%%%%%%%%%%%%%%%%%%%%%%%%%%%%%%%%%%%%%%%%%%%%%%%%%%%%%%%%%%%%%%%%%%%%%%%%%%%%%
\section{Introduction}
Quantum phase estimation (QPE) is a commonly used technique in many important algorithms, such as prime factorization\cite{Shor:05}, 
quantum walk \cite{Szegedy:04}, discrete logarithm\cite{Shor:94} and quantum counting\cite{BHT:98}. Various approaches have been 
devised to implement QPE and all have different requirements. For instance, 
the standard QFT$^\dagger$ \cite{NC:00} requires high precision of the control rotation gates. The one by Kitaev \cite{KSV:02, Kitaev:95} 
requires constant rotation gates and Hadamard gates. The advantage of the former is that it does not require repetition
while the latter needs repetition and classical post-measurement process in order to achieve the required precision. 
Between these two extreme approaches, there exist some approaches \cite{HC:11, Cheung:04, KLM:07} 
that scale between two extremes in terms of the required precision of the rotation gate. By examining the trade off 
between the rotation gate precision and the required number of repetition, those approaches tend to be of 
lower complexity.  \\

We are interested in comparing the approaches from a lower level by examining the cost that comes from the phase
kick back and the phase shift operators. Here we would like to point out that we are comparing the number of 
required elementary gates (single qubit gates and two-qubit gates). By doing so, if in the near future we can explicitly 
express the cost for obtaining a gate within certain precision, we can select the 
best (less costly) implementation based on various scenarios.  \\

For QFT$^\dagger$, Kitaev's Hadamard test is the standard and it performs
efficiently with classical post measurement processing. In \cite{HC:11}
it is shown that its circuit depth for QFT$^\dagger$ is about $1/14$ of Kitaev's approach
when the constant-precision phase shift operator is precise to the third degree.
Recently the faster phase estimation (FPE) algorithm \cite{SHF:13} shows FPE has a $\log*$ factor of reduction in terms 
of the total number of measurements in comparison to Kitaev's approach. The core of this algorithm is similar to
Kitaev's approach but it reduces the number of measurements by use of {\em multiple bit inference}. Despite of the reduction 
in the number of measurements, this algorithm still must
invoke the corresponding unitary for phase kick back for a certain number of times. We investigate the 
required number of invocations of unitary $U$ in FPE and compare with two other approaches.

In our analysis, we assume that 
\begin{itemize}
  \item A unitary $U^{m}$ is implemented by applying the unitary $U$ $m$ times.  
  \item We use the Chernoff bound to obtain the {\em minimal} number of trials needed for each approach. Based on the required 
  {\em minimal trials}, we compare the number of elementary gates used in each approach. 
\end{itemize}
The remainder of this article is organized as the following: we briefly describe
Kitaev's approach, ACPA and FPE in section \ref{sect:overview} and provide the analysis for obtaining the required repetition of
each approach. In section \ref{sect:cost} we compare the circuit complexity and discuss the scenario when the
phase shift operation is imperfect for the ACPA. 
%%%%%%%%%%%%%%%%%%%%%%%%%%%%%%%%%%%%%%%%%%%%%%%%%%%%%%%%%%%%%%%%%%%%%%%%%%%%%%%
\section{Overview and Analysis }\label{sect:overview}
There are various settings for phase estimation. For instance, based on the availability of eigenvector. If there are multiple copies, 
approaches based on Hadamard tests can be run in parallel. Therefore, in terms of time complexity, Hadamard test based approaches 
should be chosen. If there is only one copy, then time complexity of all approaches are proportional to their circuit size. Here
we consider there is only one copy of eigenvector as we are interested in the number of required elementary gates used. 

\begin{problem}{\bf [Phase Estimation]} 
Let $U$ be a unitary matrix with eigenvalue $e^{2\pi i \varphi}$ and  corresponding eigenvector $|u\>$. 
Assume only a single copy of $\ket{u}$ is available, the goal is to find $\widetilde{\varphi}$ such that
\begin{equation}
\Pr(|\widetilde{\varphi}-\varphi|<\frac{1}{2^{n}})> 1-c,
\end{equation}
where $c$ is a constant less than $\frac{1}{2}$.
\end{problem}
To simplify the analysis, let us choose $c = 1/4$ such that the result from phase estimation is precise
to the $n_{th}$ bit and the success probability is at least $3/4$. To achieve such a goal, 
in this section we will compute the numbers of trials needed for Kitaev's approach, 
ACPA and FPE.

\subsection{Kitaev's original approach}\label{sect:kitaev}

\begin{theorem}\cite{HC:11}
	Assume $U$ is a unitary matrix with eigenvalue $e^{2\pi i \varphi}$ and
	corresponding eigenvector $|u\>$. Suppose $\varphi = 0.\varphi_1\ldots
	\varphi_n $ and let $\phi_l = 2^{l-1}\varphi$ ($1\leq l\leq n$). To obtain the recovered
	$\tilde{\varphi}$ that is precise to the $n_{th}$ bit with constant success probability
	greater than $1-c$ where $c < 1/2$, for each $\phi_l$ we need to run at least $55 \ln
	\frac{n}{c}$ trials of Hadamard tests when using Kitaev's approach.
	\end{theorem}
To be complete, here we briefly describe the analysis given in \cite{HC:11} for the above theorem. 
For the interested readers, a similar analysis on the FPE is given in section \ref{sect:FPE}. In Kitaev's approach, 
a series of Hadamard tests are performed.  
In each test the phase $2^{k-1}\varphi$ ($1\leq k\leq n$) will be 
computed up to precision $1/16$. Assume an $n$-bit approximation is desired. 
Starting  from $k=n$, in each step the $k_{th}$ bit 
position is determined consistently from the results of previous steps. 

\begin{figure}[htbp]
% 					\[	\Qcircuit @C=1em @R=1em {
% 							\lstick{\ket{0}}   & \gate{H}  & \gate{K}     &\ctrl{1}       &\gate{H}  &\qw &\meter\\
% 							\lstick{\ket{u}}                                  & \qw       & \qw          &\gate{U^{2^{k-1}}} &\qw       &\qw & & \lstick{\ket{u}}   }\]
					\begin{center}
					\includegraphics[height=1.1in,width=3.7in]{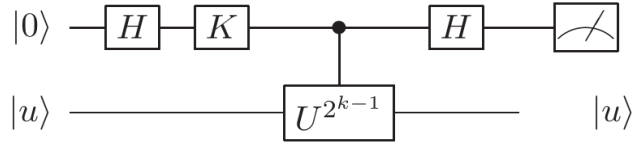} 					
					\caption{Hadamard test with extra phase shift operator.} \label{QPE_with_K_Operator}
					\end{center}
					\end{figure}

For the $k_{th}$ bit position, we perform the Hadamard test depicted in  Figure~\ref{QPE_with_K_Operator}. 
Denote $\varphi_k = 2^{k-1}\varphi$ and choose gate $K=I_2$, the probability of the post measurement state is
\begin{equation}\label{Eq:P1}
\Pr(0|k) = \frac{1 + \cos(2\pi \varphi_k)}{2}, \quad 
\Pr(1|k) = \frac{1 - \cos(2\pi \varphi_k)}{2}. 
\end{equation} 
By choosing the gate
\begin{equation}
K = \left( {\begin{array}{cc} 1 & 0  \\
 0 & i  \\
 \end{array} } \right) 
\end{equation} 
as the square root of Pauli-Z ($\sigma_z$) gate, we have 

\begin{equation}\label{Eq:P2}
\Pr(0|k)=\frac{1 - \sin(2\pi \varphi_k)}{2}, \quad 
\Pr(1|k) = \frac{1 + \sin(2\pi \varphi_k)}{2}. 
\end{equation}

From the estimates of the probabilities, we have enough information to recover $\varphi_k$. In Kitaev's original approach, 
after performing the Hadamard tests, some classical post processing is also necessary.
Suppose $\varphi = 0.x_1 x_2\ldots x_n$ is an exact $n$-bit. If we are able to
determine the values of $\varphi$, $2\varphi, \ldots ,$ $2^{n-1} \varphi$ with some
constant-precision ($1/16$ to be exact), then we can determine
$\varphi$ with precision $1/2^n$ efficiently \cite{Kitaev:95, KSV:02}.

Starting with $\varphi_n$ we increase the precision of the estimated fraction as we proceed toward
$\varphi_1$. The approximated values of $\varphi_k \,(k = n, \ldots, 1)$
will allow us to make the right choices.

For $k=1,\ldots,n$ the value of $\varphi_k$ is replaced by 
$\beta_k$, where $\beta_k$ is the closest number chosen from the
set $\{\frac{0}{8},\frac{1}{8},\frac{2}{8},\frac{3}{8},\frac{4}{8},\frac{5}{8},\frac{6}{8},\frac{7}{8} \}$ 
such that
\begin{equation}\label{eqn:pa}
|\varphi_k - \beta_k|_{\text{mod 1}} <  \frac{1}{8}. 
\end{equation}

The result follows by a simple iteration. Let $\beta_n=\overline{0.x_n x_{n+1} x_{n+2}}$ and proceed by the following iteration:

\begin{equation}
x_k = \left\{ 
\begin{array}{l l}
  0 & \quad \mbox{if \quad $|\overline{0.0x_{k+1}x_{k+2}} - \beta_{k}|_{\text{mod 1}} < 1/4$ }\\ 
  1 & \quad \mbox{if \quad $|\overline{0.1x_{k+1}x_{k+2}} - \beta_{k}|_{\text{mod 1}} < 1/4$}\\ \end{array} \right. 
\end{equation}
\noindent 
for $k= n-1, \ldots, 1$. By using simple induction, the result satisfies the following inequality:
\begin{equation}
|\overline{0.x_1x_2\ldots x_{n+2}} - \varphi|_{\text{mod 1}} < 2^{-(n+2)}. 
\end{equation}

In Eq.~\ref{eqn:pa}, we do not have the exact value of $\varphi_k$. So, we have to estimate 
this value and use the estimate to find $\beta_k$. Let $\widetilde{\varphi_k}$ be the estimated value and
\begin{equation}
\epsilon=|\widetilde{\varphi_k}-\varphi_k|_{\text{mod 1}}
\end{equation} 
be the estimation error. Now we use the estimate to find the closest $\beta_k$. Since we know the exact binary 
representation of the estimate $\widetilde{\varphi_k}$, we can choose $\beta_k$ such that
\begin{equation}
|\widetilde{\varphi_k}-\beta_k|_{\text{mod 1}}\leq\frac{1}{16}.
\end{equation} 

By the triangle inequality we have,
\begin{equation}
|\varphi_k-\beta_k|_{\text{mod 1}}\leq |\widetilde{\varphi_k}-\varphi_k| _{\text{mod 1}}+ |\widetilde{\varphi_k}-\beta_k|_{\text{mod 1}}\leq\epsilon +\frac{1}{16}.
\end{equation} 

To  satisfy Eq.~\ref{eqn:pa}, we need to have $\epsilon<1/16$, which implies
\begin{equation}\label{eqn:16}
|\widetilde{\varphi_k}-\varphi_k|_{\text{mod 1}}<\frac{1}{16}.
\end{equation} 
Therefore, it is necessary that the phase be estimated with precision $1/16$  at each stage.
Let $s_k$ be the estimate of $\sin(2\pi\varphi_k)$ and $t_k$ the estimate of $\cos(2\pi\varphi_k)$. 
By Eq.~\ref{eqn:16} we should have
\begin{equation}
\left|\varphi_k-\frac{1}{2\pi}\arctan\left(\frac{s_k}{t_k}\right)\right|_{\text{mod 1}}<\frac{1}{16}.
\end{equation}

To estimate the phase $\varphi_k$ with precision $1/16$, $s_k$ and $t_k$ must be estimated with error 
at most around 0.2706 \cite{HC:11}. By use of Chernoff bound, in order to obtain 
\begin{equation}
\mathrm{Pr}\left(|\widetilde{\varphi_k}-\varphi_k|< \frac{1}{16}\right)> 1-\varepsilon,
\end{equation} 
we require a minimum of 
\begin{equation}
m \approx 76+55\ln \frac{1}{\varepsilon}
\end{equation} 
many trials. And by use of union bound,  we desire $1 - n\varepsilon \geq 3/4$. In our case $\varepsilon = \frac{1}{4n}$, hence  
\begin{equation}
m \approx 76+55\ln 4n . 
\end{equation} 
Therefore, for this approach the number of total unitary $U$ 
invocation will be 
\begin{equation}
 (76+55\ln 4n) \times (2^n -1). 
\end{equation}

\subsection{Arbitrary Constant-Precision Approach}\label{sect:ourAlg}
\begin{figure}[h]
    	  \begin{center}
 		  \includegraphics[height=1.3in,width=3.7in]{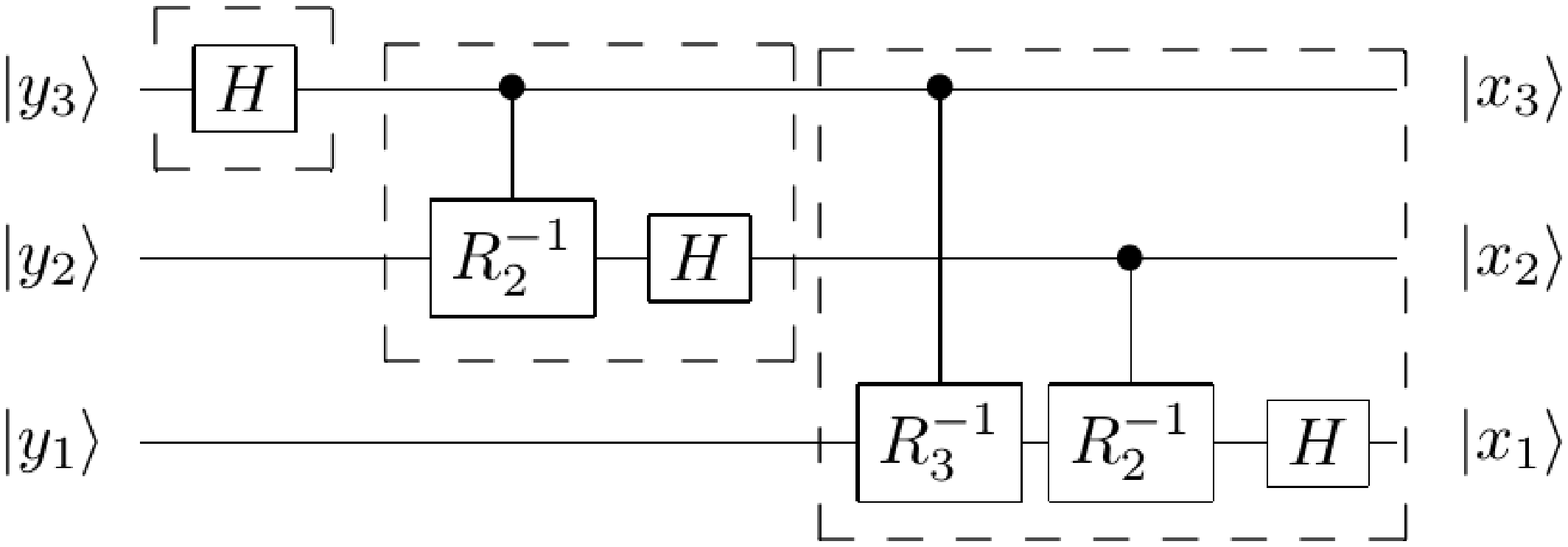} 
	      \end{center}
          \caption{3-qubit inverse QFT where $1 \leq i \leq 3$,
		  $|y_i\>=\frac{1}{\sqrt{2}}(\left|0\right>+e^{2\pi i(0.x_i\ldots x_3)}\left|1\right>$.} \label{QFT_Ours}
          \end{figure}

The arbitrary constant-precision approach (ACPA) draws a trade-off between the highest degree of phase shift operators being used 
and the depth of the circuit. As a result, when smaller degrees of phase shift operators are used, the depth of the circuit
increases and vice versa. ACPA is useful when the precision of the rotation gate is limited.  
Let rotation gate $R_k$, that is precise to the $k_{th}$ degree, be defined as follows 
\begin{equation}\label{eq:phaseshift}
						 R_k \equiv
						\left[ {\begin{array}{cc}
						 1 & 0  \\
						 0 & e^{2\pi i/2^k}  \\
						 \end{array} } \right].
\end{equation}
A simple example of $k=3$ for estimating the third least significant eigenphase bit is given in Fig.~\ref{QFT_Ours}. 

Assume the eigenphase is $\varphi=0.x_1x_2x_3\ldots x_n$ and suppose we are at step $j$, that is we are to estimate
$x_j$. It is clear that after applying controlled phase shift operators $R_2^{-1}$, $R_3^{-1}$, \ldots, $R_k^{-1}$, we obtain
 \begin{equation}\ket{\widetilde{\psi_j}}=\frac{1}{\sqrt{2}}(\ket{0}+e^{2\pi i \widetilde{\varphi}}\ket{1}),\end{equation}
where 
\begin{equation}\widetilde{\varphi}=0.x_{j}\underbrace{00\ldots 0}_\text{$k$ 0s}x_{j+k}x_{j+k+1}\ldots.\end{equation}
  
It is easy to see that
 \begin{equation}|\widetilde{\varphi}-0.x_{j}|<\frac{1}{2^k}.\end{equation}
 Hence, we can express
 \begin{equation}
 \widetilde{\varphi}=0.x_{j}+\theta. 
 \end{equation}
The post measurement probabilities of 
achieving $|0\>$ or $|1\>$ where $x_{j+1}$ = 0 \footnote{A  similar analysis can be applied to $x_{j+1}$ = 1. } are
\begin{equation}\label{correctProb}
	\Pr(0|j) = \cos^2(\pi \theta), \quad \Pr(1|j) = \sin^2(\pi \theta) 
\end{equation}
where $\theta < \frac{1}{2^k}$. Let $\tau = \frac{\pi}{2^{k-1}}$ and
we can rewrite the above formula as 
\begin{equation}\label{correctProb_1}
P(0|j) = \cos^2(\pi \theta) \geq \cos^2(\frac{\pi}{2^{k}}) =
\frac{\cos(\tau)+1}{2}.
\end{equation}

By use of Taylor expansion for the 
cosine function, we obtain a lower bound that
\begin{equation}\label{correctProb_2}
\Pr(0|j) >  1-(\tau/2)^2. 
\end{equation} 
In order to achieve a success probability of $1 -\varepsilon$ for estimating the
$j_{th}$ bit, we can bound the required number of iterations, $m$, from below. For simplicity, let us denote $p$ as $\Pr(0|j)$ and
let $X_i$ be the random variable for Bernoulli trials.  By Chernoff bound, we
have
\begin{equation}
				\mathrm{Pr}\left(\frac{1}{m}\sum_{i=0}^{m}X_i\leq \frac{1}{2}\right)\geq 1-
				e^{-2m(p-\frac{1}{2})^2 } = 1 -\varepsilon. 
\end{equation}
Hence, we would need at least
\begin{equation}\label{repeatNum}
m \geq \frac{\ln (1/\varepsilon)}{2(p-1/2)^2} 
\end{equation}
trials for estimating each eigenphase bit in order to achieve success probability
$1-\varepsilon$ (each bit) by use of rotation gates precise up to the $k_{th}$
degree. By Eq. \ref{correctProb_2}, we know that it is sufficient to choose 
\begin{equation}\label{repeatNum2}
 m = \frac{2 \ln
(1/\varepsilon)}{{(1-\frac{(\tau)^2}{2})^2}} =  \frac{2 \ln(4n)}{{(1-\frac{\pi^2}{2^{2k-1}})^2}}.
\end{equation}

Therefore, for this approach the number of unitary $U$ invocation will be 
\begin{equation}
\frac{2 \ln(4n)}{{(1-\frac{\pi^2}{2^{2k-1}})^2}} \times (2^n -1) . 
\end{equation}
For instance, when $k=3$, this approach requires about $1/14$ of trials of Kitaev's method to achieve the
same degree of success probability for each bit. However, this rough comparison is only in terms of measurements (trials). 
Kitaev's approach does not require higher degree of rotation gates, except the Hadamard. In our approach we need 
constant-precision rotation gates and this should be taken into account when comparing the cost.

\subsection{Faster Phase Estimation}\label{sect:FPE}
The Faster Phase Estimation (FPE) by Svore and et. al. \cite{SHF:13} cleverly modifies Kitaev's original approach and 
applies the two-stage-multiple-round strategy. By such modifications, FPE reduces the number of measurements by a logarithm factor, $\log^*$, 
in comparison to Kitaev's approach. However, for each measurement it uses a multiple bits inference that introduces an extra logarithm factor
in terms of the required number of invoking the unitary $U$. 

\begin{figure}[htb]
%           \[	\Qcircuit @C=1em @R=1em {
% 							\lstick{\ket{0}}   & \gate{H}  & \gate{K}     &\ctrl{1}       &\gate{H}  &\qw &\meter\\
% 							\lstick{\ket{u}}                                  & \qw       & \qw          &\gate{U^{2^{t_1-1}}+\ldots + U^{2^{t_S-1}}} &\qw       &\qw & & \lstick{\ket{u}}   }\]
          \begin{center}
          \includegraphics[height=1.1in,width=3.9in]{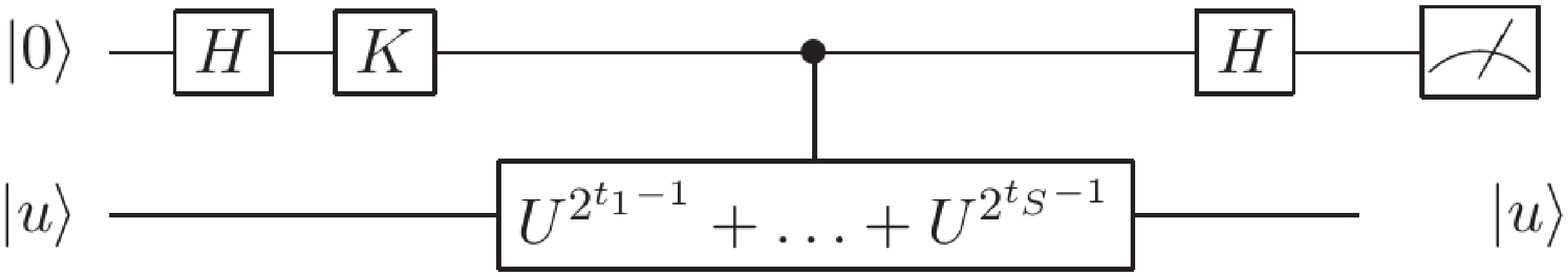} 
          \caption{FPE Stage 2: Multiple Bits Inference} \label{FPE_2}
          \end{center}
          \end{figure}

The algorithm contains two stages and the second stage contains multiple rounds. We list 
the quantum part of that algorithm in Algorithm \ref{alg:FPE} as follows:
\begin{algorithm}[H]\caption{Faster Phase Estimation}\label{alg:FPE}
First Round: \\
\textbf{for} $j = n - 1$ to $1$ \textbf{do} \\
\text{\hspace{2em} Estimate $2^{j-1}\varphi$  using $O(1)$ measurements per $j$}. \\
\textbf{end for} \\
Later Rounds: \\
\textbf{for} $r$ = 2 to Number of Rounds \textbf{do} \\
\text{\hspace{2em} Set density, $S$, and number of measurements per bit, $s_r$ , for given round.} \\
\text{\hspace{2em}\textbf{for} i = 1 to $s_r n$ \textbf{do}} \\
\text{\hspace{4em}Set $M_i$ to a sum of $S$ different powers of two, choosing these powers of two at random}\\
\text{\hspace{4em} or with a pseudo-random distribution. Perform C measurements with given $M_i$.}\\
\text{\hspace{2em}\textbf{end for}}\\
\textbf{end for} \\
Post Processing: \\
Perform multi-bit inference to determine estimate of $\beta_j =2^{j-1} \varphi$ for 
all $j$ from eigenphase estimates obtained in stage 1 and stage 2. \\
Infer eigenphase from the $\beta_j$s.  
\end{algorithm}  

The first stage (see Fig.~\ref{QPE_with_K_Operator}) is the same as Kitaev's Hadamard tests but it
requires fewer repetitions. Interested readers can refer to their work for details. The estimated eigenphase from the first stage will be used to 
obtain a more refined result in the second stage from multiple bits inference (see Fig.~\ref{FPE_2}). We refer interested
reader to \cite{SHF:13} for details. Here we are interested in finding the constant number of repetitions, referred as C in their work,
required in the second stage of FPE algorithm for resources comparison with other approaches. As shown in \cite{SHF:13}, 
it is necessary to obtain an estimate of $M_i \varphi$, let us say $\sigma_i$, with precision of $1/32$ and failure 
probability $\leq 1/8$, that is 
\begin{equation}\label{eqn:Crepeat}
\Pr[|\varphi{M_i} - \sigma_i|_{\text{mod 1}} > 1/32] < 1/8. 
\end{equation}

For simplicity, let us look at the 2-round algorithm of FPE and $M_i = \sum_{j \in S} 2^j$, $S=\{s \in \mathbb{N}|1 \leq s \leq n\}$ 
and\footnote{No matter how many rounds there are, the very last round must have S of size $\log n$} $|S| \approx \log n$ \cite{SHF:13}. 
Denote $\varphi M_i$ as $\varphi_i$. When $K=I_2$, the probability of the post measurement state is
\begin{equation}\label{Eq:P1}
\Pr(0|i)= \frac{1 + \cos(2\pi \varphi_i)}{2}, \quad 
\Pr(1|i) = \frac{1 - \cos(2\pi \varphi_i)}{2}. 
\end{equation}
In order to recover $\varphi_i$, we can estimate $\Pr(0|i)$ with higher probabilities by iterating the process. 
But we also need to distinguish between $\varphi_i$ and $-\varphi_i$. This can be solved by the same Hadamard 
test in Figure~\ref{QPE_with_K_Operator}, 
but instead we use the gate $K = \sqrt{\sigma_z}$. The probabilities of the 
post-measurement states based on the modified Hadamard test become
% \begin{figure}[h]
% \begin{center}
% 	\Qcircuit @C=1em @R=1em {
% 		\lstick{|0\>}   & \gate{H}  & \gate{K}     &\ctrl{1}       &\gate{H}  &\qw &\meter\\
% 		\lstick{|u\>}   & \qw       & \qw          &\gate{U^{2^{k-1}}} &\qw   &\qw & & \lstick{|u\>}   }
%     \caption{Hadamard test with extra phase shift operator.} \label{QPE_with_K_Operator}
% \end{center}
% \end{figure}

\begin{equation}\label{Eq:P2}
\Pr(0|i)=\frac{1 - \sin(2\pi \varphi_i)}{2}, \quad 
\Pr(1|i) = \frac{1 + \sin(2\pi \varphi_i)}{2}. 
\end{equation}
Hence, we have enough information to recover $\varphi_k$ from the estimates of the probabilities.

In the first  Hadamard test (Eq.~\ref{Eq:P1}), in order to estimate $\Pr(1|i)$ an iteration of Hadamard 
tests should be applied to obtain the required precision of $1/32$ (see Eq.~\ref{eqn:Crepeat}) 
for $\varphi_i$.  To get an estimate of $\Pr(1|i)$, we can count the number of states $\ket{1}$ in the 
post measurement state and then divide that number by the total number of iterations. 

The Hadamard test outputs $\ket{0}$ or $\ket{1}$ with a fixed probability. We can model an iteration of 
Hadamard tests as Bernoulli trials with success probability (obtaining $\ket{1}$) being $p_i$. The best 
estimate for the probability of obtaining the post measurement state $\ket{1}$   with $\kappa$ samples is
\begin{equation}\label{eqn:est1}
\widetilde{p_i}=\frac{h}{\kappa},
\end{equation} 
where $h$ is the number of ones in $\kappa$ trials. In order to find $\sin(2\pi\varphi_i)$ and $\cos(2\pi\varphi_i)$, 
we can use estimates of probabilities in Eq.~\ref{Eq:P1} and Eq.~\ref{Eq:P2}.
Let $s_i$ be the estimate of $\sin(2\pi\varphi_i)$ and $t_i$ the estimate of $\cos(2\pi\varphi_i)$. It is clear that if 
\begin{equation}
|\widetilde{p_i}-p_i|<\delta/2,
\end{equation} 
then   
\begin{equation}
|s_i-\sin(2\pi\varphi_i)|< \delta,\quad
|t_i-\cos(2\pi\varphi_i)|< \delta.
\end{equation} 

As the inverse tangent function is more robust to error than the inverse sine or cosine functions, we use 
\begin{equation}
\widetilde{\varphi_i}=\frac{1}{2\pi}\arctan\left(\frac{s_i}{t_i}\right)
\end{equation}
as the estimation of $\varphi_i$. By the condition given Eq.~(\ref{eqn:Crepeat}), the following constraint 
\begin{equation}
\Pr[|\varphi{M_i} - \frac{1}{2\pi} tan^{-1}(\frac{s_i}{t_i})| \leq 1/32]_{\text{mod 1}} > 7/8
\end{equation} 
must be satisfied.

The inverse tangent function can not distinguish between the two values $\varphi_i$ and $\varphi_i \pm 1/2$. However, 
because we find estimates of the sine and cosine functions as well, the correct value can be determined properly.
It is easy to see, in order to estimate the phase $\varphi_i$ with precision $1/32$, we should make sure that
\begin{equation}
\left|\tan^{-1}\frac{ \sin(2\pi \varphi_i)}{ \sqrt{1 - \sin^2(2\pi \varphi_i)}} - \tan^{-1}\frac{ \sin(2\pi \varphi_i) + \delta}{ \sqrt{1 - \sin^2(2\pi \varphi_i)} - \delta} \right| \leq \pi/16
\end{equation}
\noindent 
is satisfied. 
We obtain $\delta \leq \frac{1}{4}\sqrt{1-\frac{1}{\sqrt{2}}}$ and therefore the error of estimated probabilities should be 
less than $\frac{1}{8}\sqrt{1-\frac{1}{\sqrt{2}}}$. By use of the simpler bound of Chernoff-Hoeffding theorem, the number of repetition, $m$, should satisfy
\begin{equation} 
\mathrm{Pr}\left(\left|\frac{1}{m}\sum_{i=0}^{m}X_i-p_i\right| > \frac{\delta}{2}\right)\leq 2e^{-2(\frac{\delta}{2})^2 m} \leq 1/16.
\end{equation}
Because we have to estimate both sine and cosine, the failure probability is $1/16$, instead of $1/8$. The total number of 
repetition is therefore $C =2m \approx 756$. 
The algorithm later generates the estimate of each $\beta_j$ that needs to satisfy 
\begin{equation}
\Pr[|2^{j-1}\varphi - \beta_j|_\text{mod 1}> 1/16] < e^{-s_2S} \leq 1/4n.
\end{equation}
Hence, if we choose $S = \ln n$, then $s_2 = {\ln 4n}/{\ln n}$. If we are sampling from a uniform distribution for selecting
$S$ different numbers for each iteration $i$, then for $s_2n$ iterations each unitary $U^{2^{l -1}}$ where $ 1 \leq l \leq n$
will be summoned at least $s_2 \cdot S \cdot C$ times. Therefore, the number of unitary $U$ 
invocation will be $s_2\cdot S \cdot C \cdot (2^n -1)$.

\section{Cost Estimation }\label{sect:cost} 
The cost of phase estimation comes from phase kick back and inverse quantum Fourier transform. In this section, we will 
compare these three approaches under the perfect and the imperfect scenarios. Suppose unitary $U$ can
be decomposed into $\gamma$ logical gates. In the perfect scenario we assume that the unitary can be perfectly simulated and the cost of 
simulating a logical gate is one. In the imperfect case, a logical gate can only simulated imperfectly and we need to determine
how the imperfection affects the required number of repetition.

\subsection{Perfect Gate Generation}
The cost of phase kick back is in direct proportion to the number of required measurements. As shown in 
the analysis in section \ref{sect:overview}, In comparison to Kitaev's approach, even though FPE has significantly reduced the 
number of required measurements but it invokes at least $14$ times more than Kitaev's in 
invoking the unitary $U$ for phase kick back.  The cost of rotation gates only occurs in the ACPA method and the number of rotation gate
invocations is
\begin{equation}\label{eqn:rotCost}
k n \frac{2 \ln(4n)}{{(1-\frac{\pi^2}{2^{2k-1}})^2}}.
\end{equation}
In comparison to the cost of the phase kick back, for any $n$ the contribution from 
the cost of rotation gates is negligible. The logic is that the cost in Eq.~\ref{eqn:rotCost} will only carry a small 
factor for $\ln n$ with various $k$. Due to the fact that $k$ is integer and $k \geq 3$, the case
where the denominator is $0$, i.e. $k \approx 2.2$, never occurs. Therefore, the cost of logical gate invocation in phase kick back in 
ACPA is by far larger than the cost of rotation gate invocations ($kn\ln n$ v.s. $ \gamma^{2^n -1}$).

In Fig.~\ref{KitaevUS_Perfect} and Fig.~\ref{FPEUS_Perfect}, where $k \in [3,10]$ and $n \in [1,100]$, 
we plot the ratio between the numbers of unitary $U$ invocations among the three
approaches. We can see that as $k$ goes up, we should choose ACPA as the total invocation of the unitary $U$ is significantly reduced, 
in comparison to those that use Hadamard test. 
\newline
\begin{minipage}[h]{0.45\textwidth}
\begin{figure}[H]
		\begin{center}
		\includegraphics[scale=0.65]{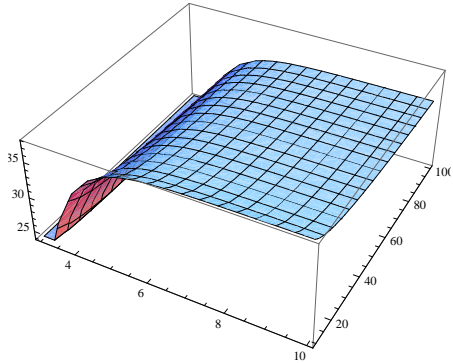}		
		\caption{Ratio: Kitaev/ours.} 
		\label{KitaevUS_Perfect}
	  	\end{center}
\end{figure}
\end{minipage}
\begin{minipage}[h]{0.10\textwidth}
\end{minipage}
\begin{minipage}[h]{0.45\textwidth}
\begin{figure}[H]
	\begin{center}
		\includegraphics[scale=0.65]{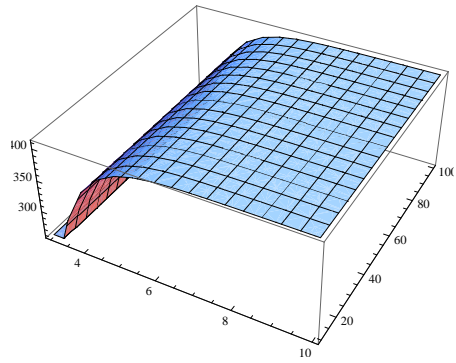}
	  	\caption{Ratio: FPE/ours.}
	  	\label{FPEUS_Perfect}
	  	\end{center}
\end{figure}
\end{minipage}

\noindent 
Here we summarize the result in Table~\ref{tbl_compare}. 

\begin{table}[h!]
\centering % used for centering table
\begin{tabular}{c c c c }  
\hline\hline  
Case & Kitaev & Faster Phase Estimation & Ours Perfect   \\ [0.5ex] % inserts table 
%heading
\hline % inserts single horizontal line
Measurements  & $ n \cdot (76 + 55\ln ({4n}))$   & $n \cdot (\log\log ({4n}) + C\cdot s_2)$ & $n \cdot \frac{2 \ln(4n)}{{(1-\frac{\pi^2}{2^{2k-1}})^2}} $  \\ % inserting body of the table
Elementary gates   &  $ (76 + 55\ln ({4n})) \cdot \gamma^{2^n -1}$ & $ (\log (\log {4n}) + S\cdot s_2\cdot C)\cdot \gamma^{2^n -1}$ & $\frac{2  \ln(4n)}{{(1-\frac{\pi^2}{2^{2k-1}})^2}}\cdot (\gamma^{2^n -1} + kn)  $\\[1ex] % [1ex] adds vertical space
\hline 
\end{tabular}
\caption{Phase kick back invocations.  $s_2 = \frac{\ln({4n})}{\ln n}, S = \ln{n}$. } % title of Table
\label{tbl_compare}
\end{table}

\subsection{Imperfect Gate Generation}
\begin{figure}[H]
		\begin{center}
		\includegraphics[scale=0.65]{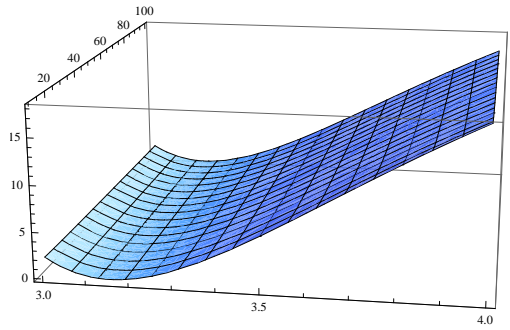}		
		\caption{Ratio: Kitaev/ours.}\label{KitaevUS_Imperfect}
	  	\end{center}
\end{figure}

In reality, we cannot generate the rotation gate perfectly. We can only generate rotation gate $\tilde R_k$ that is
$\eta$-close to rotation gate $R_k$. When considering the success probability
of estimating each eigenphase bit, the errors propagated from imperfect gate simulations must also be considered.
As mentioned in the perfect case, our approach has extra cost in the rotation gates. We need 
to examine the impact of imperfect rotation gates on the success probability because it directly 
affects the required number of repetition. Similar to the analysis in the perfect case, we have 
the same constraints but the analysis slightly varies. The Eq.~\ref{correctProb} would become  
\begin{equation}\label{correctProb_imp}
	\Pr(0|j) = \cos^2(\pi \tilde{\theta}  ), \quad \Pr(1|j) = \sin^2(\pi \tilde{\theta} ) 
\end{equation}
where $\tilde{\theta} = \theta + (k-1)\eta < \frac{1}{2^k} + (k-1)\eta$ as the error propagates from the imperfect gate simulations. Let us
choose $\eta = \frac{1}{(k-1) 2^k}$. By following the same computation logic in Eq.~\ref{correctProb_1} and Eq.~\ref{correctProb_2}, we have
the success probably of estimating the eigenphase bit correctly as
\begin{equation}
P(0|j) = \cos^2(\pi \tilde{\theta}) \geq \cos^2(\frac{\pi}{2^{k-1}})=
\frac{\cos(2 \tau)+1}{2} > 1 - \tau^2 
\end{equation}
where $\tau = \frac{\pi}{2^{k-1}}$. 
Therefore, Eq.~\ref{repeatNum2} can be rewritten as 
\begin{equation}
m = \frac{2 \ln
(1/\varepsilon)}{{(1-2(\tau)^2})^2} =  \frac{2 \ln(4n)}{{(1-\frac{\pi^2}{2^{2k-3}})^2}}
\end{equation}
in this scenario. It is clear to see that the result is almost identical to the perfect case, except the value of $k$ in 
the perfect case is now shifted to the left by 1 in the imperfect case. 
From Fig.~\ref{KitaevUS_Imperfect}\footnote{As $k$ gets larger, the figure would be identical to Fig.~\ref{KitaevUS_Perfect}} 
we know that the smallest ratio occurs when $3 < k <4$ (to be exact, it should be around $3.2$ as the imperfect case 
is the perfect case shifted to the left by 1). When $k =3$, the ratio is about $1.7$ and when $k = 4$, the 
ratio is around $14$ . Because $k$ must be an integer in our case, we know that in the imperfect case
the number of required repetition is still smaller than other the two approaches. Hence, the number of unitary $U$ invocations 
in our approach still requires fewer resources. 

There are some clever approaches \cite{NC:00, DS:13} for simulating rotation gates. For instance, in a recent paper \cite{DS:13}, 
the authors show the cost of such a gate is $O(\log^c (1/\eta))$ where $1.12 \leq c \leq 2.27$ and $\eta$ is the rotation angle error.
In Solovay-Kitaev's decomposition theorem, $c$ is around $3.94$ and is bounded from below by $1$.  By applying the cost function given 
in the decomposition theorem, the extra cost in ACPA would be
\begin{equation}
kn \cdot \frac{2 \ln(4n)}{{(1-\frac{\pi^2}{2^{2k-3}})^2}}\cdot O(\log^c(1/\eta)) = O((k+ \log k)^c( kn \cdot \frac{2 \ln(4n)}{{(1-\frac{\pi^2}{2^{2k-3}})^2}})).
\end{equation}
Similar to the argument in the perfect case, the extra cost ($O((k+ \log k)^c( kn \ln n))$ v.s. $\gamma^{2^n -1}$) is negligible 
(by use of L'H\^{o}pital's rule) when $n$ is large and $k$ is small (as we are discussing constant precision rotation gates) and $c$ is a constant less than 4. 
\section{Discussion}\label{discuss}
		In this work, we are analyzing the three approaches solely in terms of the circuit complexity.
		Hence, it pays off if higher degree of rotation gates can be implemented as it significantly
		reduces the cost from phase kick back. However, if we need to consider time complexity 
		when multiple eigenvectors are available and higher degrees of rotation gates are unfeasible,
		Hadamard test based approaches should be chosen as they can be run in parallel while the 
		ACPA can be run partially in parallel. 
		
%%%%%%%%%%%%%%%%%%%%%%%%%%%%%%%%%%%%%%%%%%%%%%%%%%%%%%%%%%%%%%%%%%%%%%%%%%%%%%%
\section{Acknowledgments}
		C.~C gratefully acknowledges the support of Lockheed Martin Corporation. We also like to thank J.~Anderson, 
		P.~Iyer and D.~Poulin for useful comments and suggestions. 
%%%%%%%%%%%%%%%%%%%%%%%%%%%%%%%%%%%%%%%%%%%%%%%%%%%%%%%%%%%%%%%%%%%%%%%%%%%%%%%

%\include{Bib}
%%%%%%%%%%%%%%%%%%%%%%%%%%%%%%%%%%%%%%%%%%%%%%
\end{document}